\documentstyle[aps,preprint]{revtex}
\begin{document}
\draft
\title{ Vortices in Schwinger-Boson Mean-Field Theory of
Two-Dimensional Quantum Antiferromagnets }
\author{Tai-Kai Ng}
\address{ department of Physics, Hong Kong University of Science and
           Technology,\\
 Clear Water Bay Road, Kowloon, Hong Kong }
\date{ \today }
\maketitle
\begin{abstract}
In this paper we study the properties of vortices in two
dimensional quantum antiferromagnets with spin magmitude $S$
on a square lattice within the framework
of Schwinger-boson mean-field theory. Based on a continuum
description, we show that vortices are stable topological excitations
in the disordered state of quantum antiferromagnets. Furthermore,
we argue that vortices can be divided into two kinds:
the first kind always carries zero angular momentum and are bosons,
whereas the second kind carries angular momentum $S$ under favourable
conditions and are fermions if $S$ is half-integer. A plausible
consequence of our results relating to the RVB theories of
High-$T_{c}$ superconductors is pointed out.
\end{abstract}
\narrowtext

\section{Introduction}
  In the past few years there has been a lot of interests in the study
of quantum antiferromagnets in two dimension on a square lattice, stimulated
by the discovery of High-$T_{c}$ superconductors\cite{an,nn}.  Among
others, one approximated approach to the quantum antiferromagnet
problems is the Schwinger-Boson Mean-field Theory (SBMFT).\cite{nn}.
In this approach, the quantum spins are represented as Schwinger bosons
and an approximated ground state is constructed by bose condensation
of spin-singlet pairs. The mean-field theory can also be formulated in a
large-N expansion as the saddle point solution to a generalized $SU(N)$
quantum spin model\cite{nn,rs}. In the limit $N \rightarrow \infty$,
the mean-field theory becomes exact. SBMFT predicts that in one dimension,
Heisenberg antiferromagnet is always disordered, with a non-zero spin
gap in the excitation spectrum. The theory is found to give an adequate
description for integer spin chains, while it fails to describe the
massless state of half-integer spin chains. This is because the
topological Berry phase term which plays crucial role in the latter
case is not taken into account correctly in SBMFT.\cite{to} However, in
two dimension topological term does not exist and SBMFT is more reliable.
In particular, the theory predicts that the disordered (spin gap) phase
exists only when the spin magnitude $S$ is small enough, when
$S<S_{c}\sim0.19$. It turns out that SBMFT offers a fairly accurate
description for the magnetic properties of the High-$T_{c}$ compounds
in the low doping regime\cite{e1}. However, at large doping levels,
the generalization of SBMFT which includes holes as fermions
\cite{g5}, does not seem to describe the High-$T_{c}$ compounds correctly.
For example, photo-emission experiments\cite{shen} indicate that
the compounds have "large" Fermi-surfaces which satisfy Luttinger
theorem, whereas the generalization of SBMFT to include holes
predicts a "small" Fermi surface with area proportional to
concentration of holes. In the large doping regime,
it turns out that the slave-boson mean-field theory\cite{sb}
which treats the spins as fermions and holes as bosons
produces a lot of properties of High-$T_{c}$
compounds correctly\cite{nl} and the SBMFT is only superier in
describing the low-doping, antiferromagnetic state. Thus a
natural question is: what is the relation between the two theories?
Can one understand the two theories within a single framework?
How does nature crossover from one description to the other as
concentration of holes increases?

   In SBMFT, the spin operator $\vec{S}_{i}$ on site $i$ is expressed
in terms of Schwinger bosons $\vec{S}_{i}=\bar{Z}_{i}\vec{\sigma}Z_{i}$,
where $\vec{\sigma}$ is the Pauli matrix, $Z_{i}(\bar{Z}_{i})$
are two component spinors $\bar{Z}_{i}=(\bar{Z}_{i\uparrow},
\bar{Z}_{i\downarrow})$, etc. Notice that in order
to represent a spin with magnitude $S$, there should be $2S$ bosons
per site\cite{nn}. The Hamiltonian can then be represented in terms
of Schwinger bosons, and a mean-field theory can be formulated
by introducing order parameters $\Delta_{i,i+\eta}=<Z_{i\uparrow}
Z_{i+\eta\downarrow}-Z_{i\downarrow}Z_{i+\eta\uparrow}> (\eta =
\pm\hat{x},\pm\hat{y})$\cite{nn}.
Alternatively, SBMFT can also be formulated in a large-N
expansion as the saddle point solution to a generalized $SU(N)$
quantum spin model. To derive the large-N theory, we divide the
square lattice into $A$ and $B$ sublattices and  consider the
following transformation for the Schwinger bosons on $B$-sublattice,
\[
Z^{B}_{j\uparrow}(\bar{Z}^{B}_{j\uparrow}) \rightarrow -Z^{B}_{j\downarrow}
(\bar{Z}^{B}_{j\downarrow}),\]
\[
Z^{B}_{j\downarrow}(\bar{Z}^{B}_{j\downarrow}) \rightarrow
Z^{B}_{j\uparrow}(\bar{Z}^{B}_{j\uparrow}),\]

for all sites $j$ on the $B$ sublattice. The Schwinger bosons on the
$A$ sublattice remains unchanged in the above transformation. The Lagrangian
of the Heisenberg model can then be represented in the transformed boson
coordinates as\cite{rs}

\begin{eqnarray}
{\em L}=\sum_{i\sigma}[\bar{Z}^{A}_{i\sigma}({d \over d\tau}+i\lambda_{i})
Z^{A}_{i\sigma}-i2S\lambda_{i}] + \sum_{j\sigma}[\bar{Z}^{B}_{j\sigma}
({d \over d\tau}+i\lambda_{j})Z^{B}_{j\sigma}-i2S\lambda_{j}]
\nonumber \\
+J\sum_{i,\eta=\pm\hat{x},\pm\hat{y}}|\Delta_{i,i+\eta}|^2 -
J\sum_{i\sigma,\eta=\pm\hat{x},\pm\hat{y}}[\Delta^{*}_{i,i+\eta}Z^{A}_{i\sigma}
Z^{B}_{i+\eta\sigma}+H.C.]
\label{lag}
\end{eqnarray}

where $\lambda_{i}$'s are Lagrange multiplier fields enforcing the constraint
that there are $2S$ bosons per site, $\Delta_{i,i+\eta}$'s are Hubbard-
Stratonovich fields introduced in decoupling of $H$. $\sigma$ is the spin
index. A $SU(N)$ spin theory can be formulated with the above Lagrangian if
the usual $SU(2)$ spins $\uparrow$ and $\downarrow$ are extented to the $SU(N)$
case where N-spin species are introduced. SBMFT can be considered as the
saddle point solution of the path integral over ${\em L}$ which becomes exact
in
the limit $N\rightarrow\infty$\cite{rs}. For infinite system with no defects,
the saddle point solution has position independent $\Delta_{i,i+\eta}$'s and
$\lambda$'s and the solution is formally very similar to the BCS solution for
superconductors, except that in present case the pairing objects are Schwinger
bosons but not electrons. Another important difference is that in the present
case, the pairing bosons are always located on different sublattices, whereas
no such distinction is found in the case of superconductors.

   The similarity between SBMFT and BCS theory leads us to ask the natural
question of whether vortex-like solutions can be found in SBMFT as in the case
of superconductors. More precisely, one may ask the question of whether one
can construct a stable solution in SBMFT where $\Delta_{i,i+\eta}$
has a phase structure
\[
\Delta_{i,i+\eta}\sim|\Delta_{i,i+\eta}|e^{i\theta(i+\eta/2)}, \]

where $\theta(\vec{x})$ is a smooth function of $\vec{x}$ and has a singular
point at $\vec{x'}\sim{0}$ such that for distance
$|\vec{x}|>>0$, $\theta(\vec{x})\rightarrow \theta(\vec{x})+2\pi$
if one moves vector $\vec{x}$ around a close loop $C$ which enclosed
the point $\vec{x'}=0$. As in the case of superconductors, the simplest
way to address this question is to construct a continuum description
for SBMFT (analogous to Ginsburg-Landau theory for superconductors)
and study the possibility of having vortex solution in the continuum
approximation. In the following we shall perform such a study in the
disordered state in SBMFT where the Schwinger bosons
are not bose-condensed and spin is a good quantum number.
In realistic High-$T_{c}$ cuprates long range antiferromagnetic
order is destroyed by introduction of charge carriers (holes). We
shall not consider complications introduced by holes here and shall
assume simply a disordered magnetic state with realistic spin
magnitudes which can be described by the spin gap phase
of SBMFT. We shall examine the properties of vortices within
the continuum description. In section II we shall derive
the continuum theory.
Based on the continuum equations, we shall first study the
properties of a single, non-magnetic impurity in the disordered
state of SBMFT, where we shall point out the important
difference between perturbations which are symmetric to the
A and B sublattices, and perturbations which distinguish the two
sublattices. In section III, we shall study static vortex excitations
in our model. Based on the continuum theory, we shall argue
that vortices are stable topological excitations in the disordered
state of 2D quantum antiferromagnets. We shall then show
that there exists two kind of vortices in the continuum theory,
corresponding to vortices centered on the mid-point of a
plaquette, and vortices centered on a lattice point. We shall
show that the properties of these two kinds of vortices are
very different, because of the different symmetry with
respect to the two sublattices. The first kind of vortex which is
symmetric to the two sublattices can carry only zero angular
momentum and is a boson, whereas the second kind of vortex
distinguishes between the two sublattices can carry angular
momentum $S$ under favourble conditions and is a fermion
if $S$ is half-integer.  In section IV we shall concentrate
ourselves at the case $S=1/2$ which is the physical case of interests
and shall examine properties of a "liquid" of fermionic vortices.
Based on simple symmetry arguments, we shall argue that the effective
theory of a liquid of fermionic vortices has precisely the same form
as the slave-boson mean-field theory for spins in the undoped limit.
The case with finite concentration of holes will also be discussed.
The findings in this paper will be summarized in section V, where
we shall discuss a plausible scenerio of how the system
crossover from a state described by SBMFT to a state described
by slave-boson mean field theory upon doping.

\section{Continuum description for SBMFT}
   To derive the continuum theory we following Read and
Sachdev\cite{rs} and consider the representation where our
system has two sites per unit cell and introduce the 'uniform' and
'staggered' components in the $\Delta$ and $\lambda$ fields, where

\begin{mathletters}
\label{fluct}
\begin{equation}
\Delta_{i,i\pm\eta}={1 \over 2}[\phi(i\pm{\eta \over 2})+q_{\pm\eta}(i\pm{\eta
\over 2})]e^{i[\theta(i\pm{\eta \over 2})+A_{\pm\eta}(i\pm{\eta \over 2})]},
\label{fl1}
\end{equation}
where $q_{-\eta}=-q_{\eta}$, $A_{-\eta}=-A_{\eta}$ and in momentum space
(for $\vec{k}\neq{0}$)
\begin{equation}
\theta_{\tau}(\vec{k})={1 \over 2}[\lambda^{A}(\vec{k})+\lambda^{B}(\vec{k})],
\label{fl2}
\end{equation}
\begin{equation}
A_{\tau}(\vec{k})={1 \over 2}[\lambda^{A}(\vec{k})-\lambda^{B}(\vec{k})].
\label{fl3}
\end{equation}
\end{mathletters}

   The $\phi(\vec{x})$ field describes uniform antiferromagnetic correlations
whereas $q(\vec{x})$ field describes spin dimerization (spin-Peierls) effects
\cite{rs}. $\theta(\vec{x})$ and $A_{\eta}(\vec{x})$ are fields describing
the corresponding uniform and staggered $phase$ fluctuations, respectively.
All fields are slowly varying on the scale of a unit cell. Notice that the
reason why four independent (real) fields are needed to describe fluctuations
in
one unit cell is precisely because we have divided the system into $A$ and $B$
sublattices. In the case of usual superconductors such a distinction is not
present and only two real fields (or one complex scalar field) enters into the
Ginsburg-Landau equation. Notice that $V_{\mu}(\vec{x})=\partial_{\mu}
\theta(\vec{x})$ and $A_{\mu}(\vec{x})$ can be considered
as $U(1)$ gauge fields coupling to the Schwinger boson $Z$'s. The
'uniform' gauge field $V_{\mu}(\vec{x})$ couples to the bosons on the
two sublattices with {\em same} gauge charge, whereas the gauge charges
for the 'staggered' gauge field $A_{\mu}(\vec{x})$ are {\em opposite}
on the two sublattices. A single vortex solution centered at $\vec{x'}=0$
corresponds to a solution of SBMFT with boundary condition
\[
\oint{V_{\mu}(\vec{x})}dx_{\mu}=2n\pi, \]

or in terms of the gauge field, there is a 'uniform' gauge flux of $n/2$ flux
quantum passing through the origin, similar to the case of superconductors.
Notice that in SBMFT where $<\Delta_{i,i+\eta}>\neq{0}$, the gauge symmetry
associated with the 'uniform' gauge field $V_{\mu}(\vec{x})$
($Z^{A}\rightarrow{Z}^{A}e^{i\Gamma},Z^{B}\rightarrow{Z}^{B}e^{i\Gamma},
V_{\mu}\rightarrow{V_{\mu}}-2\partial_{\mu}\Gamma$) is broken, whereas
the 'staggered' gauge symmetry ($Z^{A}\rightarrow{Z}^{A}e^{i\Gamma},
Z^{B}\rightarrow{Z}^{B}e^{-i\Gamma},A_{\mu}\rightarrow{A}_{\mu}-\partial_{\mu}\Gamma$)
remains intact in SBMFT. The existence of vortex solution is tied with
the broken symmetry gauge field $V_{\mu}(\vec{x})$ as in usual
superconductors.

   At distance much larger than lattice spacing,
fluctuations associated with the 'uniform' variables $\phi$
and $\theta_{\tau}$ are unimportant. Thus we shall neglect the
$\nabla\phi(x)$ and $\nabla\theta_{\tau}(x)$ terms in the following
and derive a continuum theory for the rest of the variables. Notice
that the $V_{\mu}$ variable is kept in our continuum
theory since the term is essential for studying of vortices. In the
continuum limit, the Lagrangian $L$ becomes

\begin{eqnarray}
\lefteqn{{\em L} \rightarrow }\nonumber \\
 & & \int{d\tau}\int{d^{2}x}\left\{ \sum_{\sigma}
\left\{ \right.\bar{Z}^{A}_{\sigma}(x){\partial \over \partial\tau}
Z^{A}_{\sigma}(x)+\bar{Z}^{B}_{\sigma}(x)
{\partial \over \partial\tau}Z^{B}_{\sigma}(x)\right.
\nonumber \\
 & & -\sum_{\mu=\hat{x},\hat{y}}\left[ \phi(x)(1-{1 \over 8}
(V_{\mu}(x)V^{\mu}(x)))\bar{Z}^{A}_{\sigma}(x)\bar{Z}^{B}_{\sigma}(x)-
{1 \over 2}\phi(x)(D_{\mu}\bar{Z}^{A}_{\sigma}(x))
(D_{\mu}^{*}\bar{Z}^{B}_{\sigma}(x))\right.
\nonumber \\
 & & +\left.{1 \over 2}q_{\mu}[\bar{Z}^{B}_{\sigma}(x)D_{\mu}
\bar{Z}^{A}_{\sigma}(x)-\bar{Z}^{A}_{\sigma}(x)D_{\mu}^{*}
\bar{Z}^{B}_{\sigma}(x)] + H.C.  \right]
\nonumber \\
 & & +\theta_{\tau}(x)(\bar{Z}^{A}_{\sigma}(x)
Z^{A}_{\sigma}(x)+\bar{Z}^{B}_{\sigma}(x)Z^{B}_{\sigma}(x))
\nonumber \\
 & & +\left.\left.A_{\tau}(x)(\bar{Z}^{A}_{\sigma}(x)
Z^{A}_{\sigma}(x)-\bar{Z}^{B}_{\sigma}(x)Z^{B}_{\sigma}(x)) \right\}
+{1 \over 2}\sum_{\mu}[\phi(x)^{2}+q_{\mu}(x)^{2}]-4S\theta_{\tau}(x)
\right\},
\label{lancon}
\end{eqnarray}

   where $D_{\mu}=\partial_{\mu}+iA_{\mu}$ and $D_{\mu}^{*}=
\partial_{\mu}-iA_{\mu}$. The continuum Lagrangian Eq.\ (\ref{lancon})
can be further simplified by introducing new boson fields
\[
\psi_{\sigma}={1 \over 2}(Z^{A}_{\sigma}+\bar{Z}^{B}_{\sigma}) \]
\[
\pi_{\sigma}={1 \over 2}(Z^{A}_{\sigma}-\bar{Z}^{B}_{\sigma}). \]

   The $\pi_{\sigma}$ field can be integrated out safely at large
distance and low energy\cite{rs},
leaving effective Lagrangian for $\psi_{\sigma}$ field,

\begin{eqnarray}
\lefteqn{{\em L} = } \nonumber \\
 & & \sum_{\sigma} \int{d\tau}\int{d^{2}x}\left\{ \phi(x)\left(
|D_{\mu}^{*}\psi_{\sigma}|^{2}+m(x)^{2}|\psi_{\sigma}|^{2}\right)+
(\phi_{1}(x))^{-1}(D_{\tau}-q_{\mu}D_{\mu})\psi_{\sigma}^{+}(D_{\tau}^{*}+q_{\mu}
D_{\mu}^{*})\psi_{\sigma}\right.
\nonumber \\
 & & \left. +{1 \over 2}[\phi(x)^{2}+q_{\mu}(x)^{2}]-4S \left( 2\phi(x)
(1-{1 \over 8}(V_{\mu}(x)V^{\mu}(x)))+m(x)^{2} \right) \right\}
\label{lanpsi}
\end{eqnarray}

where $\phi(\vec{x})m(\vec{x})^{2}=2\theta_{\tau}(\vec{x})-
4[1-(1/8)(V_{\mu}(\vec{x})V^{\mu}(\vec{x})]\phi(\vec{x})$ and
and $\phi_{1}(\vec{x})=(1/2)\{\theta_{\tau}(\vec{x})+
2[1-(1/8)(V_{\mu}(\vec{x})V^{\mu}(\vec{x})]\phi(\vec{x})\}$.
In the limit $q_{\mu}=0$ and $A_{\mu}=0$, and neglecting
gradient terms $\nabla\phi,\nabla\theta_{\tau}$,
the above Lagrangian can be easily diagonalized
resulting in an effective Lagrangian ${\em L_{eff}}$ in terms of
the $\phi(\vec{x})$ and $V_{\mu}(\vec{x})$
fields. The dynamical effect of the remaining terms can be obtained
by looking at Gaussian fluctuations of the fields around the saddle
point solution $A_{\mu},q_{\mu}=0$.\cite{rs} In particular, in the
small $m(\vec{x})\sim{m}\rightarrow0$ limit, we obtain (see Appendix)

\begin{eqnarray}
\lefteqn{\em L_{eff} = }  \nonumber \\
& & \int{d\tau}\int{d^{2}x} \left\{ (a-4(2S+1))\phi +
\phi^{2}+({1 \over 2}-{b \over \phi})q_{\mu}q^{\mu}
+{(2S+1)\phi \over 2}(V_{\mu}V^{\mu}) \right. \nonumber \\
 & & \left. +{1 \over 2e^{2}}
F_{\mu\nu}^{2}+icF_{\mu\tau}q_{\mu} + O(m^{2}) \right\},
\label{gl}
\end{eqnarray}

  where $e^{2}\sim{m}$, $a(<4(2S+1)),b$ and $c$ are constants of order O(1).
The precise values of $a,b,c$ depend on the underlying lattice structure
and cannot be obtained in a continuum theory. We have consider the
realistic case $N=2$ in deriving the above expression. Notice that
similar effective Lagrangian has been obtained by Read and
Sachdev previously in studying effect of instantons\cite{rs}. The
only difference here is that the $V_{\mu}V^{\mu}$ term which was not
considered by Read and Sachdev is now retained. Recall that $V_{\mu}$
is the 'uniform' $U(1)$ gauge field coupling to the
bosons and the $V_{\mu}V^{\mu}$ term in Eq.\ (\ref{gl}) just
represents the Meissner effect associated with nonzero order
parameter $\Delta$ in SBMFT. Notice that as in case of usual
superconductors, we have choosen the London gauge
$\nabla.\vec{V}=0$ in our derivation. The term can also be
written in a gauge invariant way by replacing $V_{\mu}$ by the
gauge invariant object $V_{\mu}-2\partial_{\mu}\Gamma$ in Eq.\
(\ref{gl}). This term does not contribute to the instanton
effect discussed by Read and Sachdev\cite{rs} but is
crucial to the study of vortices.

   To understand more bout the continuum theory we first consider the
properties of a single non-magnetic impurity in the disordered state
of SBMFT.\cite{nng} We shall assume that a non-magnetic impurity
simply replaces a spin at site $i$ by a non-magnetic object.
The simplest way to model this in our formalism is to is replace
the constraint equation $\bar{Z}Z=2S$ on site $i$ by
$\bar{Z}Z=2S(1-n_{i})=0$, where $n_{i}$ is the number of
non-magnetic impurity on site $i$. It is than easy to see that the
effect of nonzero $n_{i}$ in our effective Lagrangian Eq.\ (\ref{gl})
is to introduce in the Coulomb gauge an extra term
$-(2Sn_{i}e_{i})A_{\tau}$, where
$e_{i}=\pm{1}$, depending on whether the impurity is located on
the $A$- or $B$- sublattices,\cite{nng} i.e., non-magnetic
impurities appear as effective (staggered) gauge charges of
magnitude $2Se$ localized at the impurity sites. A corresponding
electrostatic potential $V(r)\sim(2Se_{i})ln(r/\xi)$
(where $\xi\sim{m}^{-1}$) will be induced around the impurity.
To lower the electrostatic energy, $2S$ bosons will be nucleated
out from the vacuum to screen the electric field, resulting in
formation of local magnetic moment of magnitude $S$ around
a non-magnetic impurity\cite{nng}. Notice that this effect has
been observed experimentally in the High-$T_{c}$ compounds in the
underdoped region upon substitution of Cu by Zn in the
conducting planes\cite{nm}.

   This result can also be understood in an alternative way by
observing that the same physical effect should have occurred if
instead of replacing a spin on site $i$ by an non-magnetic
impurity, we let the Heisenberg coupling $J_i$ to go to zero for
those bonds joining to site $i$. In this case the 'impurity' site
$i$ behaves as an non-magnetic impurity as far as the rest of
the system is concerned. The only difference is that a free
spin of magnitude $S$ now remains on site $i$.
However, the effect of this approach on our effective Lagrangian
Eq.\ (\ref{gl}) looks rather different compared with our previous
approach. Instead of introducing a 'staggered' electric charge
of magnitude $2Se$ on site $i$, a boundary condition
$\Delta_{i,i+\eta}=0 (\eta=\pm\hat{x},\pm\hat{y})$ is now imposed
on our system. Using Eq.\ (\ref{fl1}), the corresponding boundary
condition in the continuum theory becomes
\[
q_{\hat{n}}(x_{i})=\pm\hat{n}\phi(x_{i})
\]
  where $i\rightarrow{x_{i}}$ in the continuum limit. $\hat{n}$ is
a unit vector. $\pm$ depends on whether site $i$ is on $A$- or
$B$- sublattices. The properties of the system
around the non-magnetic impurity is obtained by minimizing the
free energy Eq.\ (\ref{gl}) under this boundary condition.

   Performing the calculation, we find first of all that
$\phi(x_{i})\neq0$ and correspondingly $\vec{q}(x_{i})\neq0$. The
reason for this behaviour can be understood easily from Eq.\ (\ref{gl})
by noticing that the coefficient in front of the $q_{\mu}q^{\mu}$ term
becomes negative as $\phi$ becomes smaller than $2b$, implying that
spontaneously dimerization will occur when $\phi$ becomes small enough.
Because of stability requirement, it can be shown that dimmerization
$|\vec{q}|$ cannot have value larger than $q_{o}=(\surd{2})\phi$
(see Appendix).
Anyway both $\phi(x_{i})$ and $\vec{q}(x_{i})\sim\vec{q_{o}}$
should be nonzero in the continuum solution.
The reason why $\vec{q}$ should be nonzero can also be understood
from symmetry consideration. Notice that the $\phi(x)$ field is
symmetric under exchange of $A$- and $B$- sublattices whereas
$q_{\mu}(x)$ field is antisymmetric.
Thus a solution with $q_{\mu}(x)$ field identically zero would not distinguish
between the two sublattices. However, the single non-magnetic impurity
problem certainly distinguishes the two sublattices since the impurity
can only be placed on either one of the two sublattices. Thus we
expect $q_{\mu}(x)$ to be nonzero around a single impurity.
Minimizing the free energy with respect to the $q_{\mu}$ field
we find also that $\vec{E}(x)\sim\vec{q}(x)$, where $E_{\mu}=i
F_{\mu\tau}$ is the 'staggered' electric field in the system. With the
boundary condition $\vec{q}\sim(\pm\hat{n})\phi$ around the impurity
we find that electric field radiating out (or in) from the impurity is
obtained in our solution, i.e., depending on whether the impurity
is located on the $A$- or $B$- sublattices, it behaves as source or
sink for the electric field, in exact agreement with our previous
approach which predicts that non-magnetic impurity behaves as a single
staggered gauge charge added to the system.

   One may also extend our approach to discuss the case when the
Heisenberg coupling $J'$ joining to the site $i$ is small but nonzero,
so that the spin at site $i$ still couples weakly with the
surrounding environment. The previous discussions still apply, except
that the spin at site $i$ appears now as a localized spinon state
occupied by $2S$ spinons in the system (notice that at two
dimension, an arbitrary small attractive interaction is enough
to generate a bound state\cite{bd}). The spinon state again behaves as a
(staggered) gauge charge (with magnitude $2S$), and will bind $2S$
other spinons with opposite gauge charge (or localized on opposite
sublattice) next to itself as in the case with $J'=0$. However for
nonzero $J'$ the localized spinons on the two sublattices will
interact antiferromagnetically because of the underlying
antiferromagnetic correlation in the system,\cite{ik}
forming at the end a localized
spin singlet around site $i$. It is only in the limit
$J'\rightarrow0$ that the two spinons are decoupled and free
local magnetic moments appear.

   Our discussion can also be extended to
study other forms of local defects. For example, one may study
the situation when the Heisenberg coupling $J_{i,i+\hat{x}}$ between
sites $i$ and $i+\hat{x}$ is set to zero,
i.e., a single bond is being removed from the system. Notice that
the perturbation is {\em symmetric} with respect to interchange of
$A$- and $B$- sublattices. It is straightforward to generalize
our previous discussion to this case where we find that the
perturbation now appears as an effective electric dipole moment
in ${\em L_{eff}}$. No local moment is expected to form around the
impurity bond in this case because of the follow reason: because
of the symmetry between $A$- and $B$- sublattices bosons(spinons) can
be nucleated out from vaccuum only in pairs, with one localized on
the $A$ sublattice and one on the $B$ sublattice. The interaction
between the nucleated spinon pairs will again be antiferromagnetic and
has a magnitude $\sim{J}e^{-x/\xi}$, where $x$ is the distance
between the two spinons and $\xi\sim{m^{-1}}$ is the 'size' of
the spinon wavefunction. For the single bond defect, the distance
$x$ between the two spinons is of order of lattice spacing $<<\xi$.
Thus the effective interaction between the two spinons is of order
$J$, i.e., they will form a strong local spin singlet and no
isolated magnetic moment will appear.

    Experimentally, the Cu ion in the conducting plane of high-$T_{c}$
cupartes can be substituted by Zn (non-magnetic impurity)\cite{nm}
and magnitude of Heisenberg exchange $J$ can be modified locally
by introducing impurities out of conduction
planes or substitution of O ions. It is observed that while
substitution of Cu by Zn in the conducting plane introduces
local magnetic moments in the underdoped (spin gap) phases of
High-$T_{c}$ cuprates\cite{nm}, no such effect is observed
in other ways of introducing defects. Our theoretical investigation
on disordered state of quantum antiferromagnets based on
continuum description of SBMFT is in satisfying
agreement with experimental results.\cite{nng}

\section{Static Vortices in disordered state of SBMFT}

   In the continuum theory, a unit vortex centered as $\vec{x}=0$
corresponds to a solution of
$\delta{\em L_{eff}}=0$, with $\theta(r,\Omega)=\Omega$,
where $r$ and $\Omega$ are the distance and angle in the
(2D) polar co-ordinate. The nonzero vorticity introduces a diverging
kinetic energy through the $V_{\mu}V^{\mu}\sim(\nabla\theta)^{2}$
term in ${\em L_{eff}}$,
which supresses the magnitude of $\phi$ around the vortex core
$r\rightarrow0$, as in the case of usual superconductors. Minimizing
the free energy with respect to $\phi$, and keeping in mind that for
small $\phi$, $|\vec{q}|\sim(\surd{2})\phi$, we find that
$\phi=|\vec{q}|=0$ for $r<r_{o}\sim[(8(2S+1)+4b-2a)/(2S+1)]^{1/2}$,
but not going to zero smoothly as
$r\rightarrow0$, as in the case of usual superconductors. This
result is an artifact of our continuum theory where the
$(\nabla\phi)^{2}$ term is not included in ${\em L_{eff}}$. Nevertheless,
the qualitative effect where $\phi$ is being suppressed around the
vortex core is clear. For large $r$, $\phi(r)\rightarrow\phi_o=
(4(2S+1)-a)/2$ and $q\rightarrow0$, indicating that vortices are stable
topological excitations in SBMFT. The behaviour of vortices in SBMFT
is very similar to vortices in usual superconductors at large
$r$. However, the small $r$ behaviours are very different.
The suppression of $\phi$ and $q$ fields around
vortex core reflects the fact that around the center of vortex, the
bond amplitude $\Delta_{i,i+\mu}$'s are being suppressed. As has
been discussed in the previous section, suppression of local bond
amplitudes may lead to generation of localized magnetic moments,
depending on the detailed bond configuration. To understand the
properties of a vortex, it is thus important to understand the
underlying bond structure at the vortex core.

   Before studying the bond structure at vortex core, let us explain
first how local magnetic moment binding to vortex core modifies the
properties of a vortex. For a unit vortex the magnetic flux
seen by a Schwinger boson is half flux quantum.
(recall that $V_{\mu}$ is an 'uniform' gauge field which has
same gauge charge for bosons on both sublattices) Therefore the
orbital angular momentum of a Schwinger boson around a unit
vortex is ${1 \over 2} + integer$. The composite object of {\em one}
Schwinger boson bound to the vortex is thus a {\em fermion} because
of the angular-momentum-statistics relation.\cite{wil} Notice that this
phenomenon has a direct correspondence in superconductors where
the bound state of an electron to a vortex is a boson.\cite{wil}
However, if instead of one boson, {\em two} bosons are
bound to a vortex, then the composite object is again a boson since
angular momentum will be of integer value again. Now it should be
clear why presence of local magnetic moment at vortex core
is an importance issue. The magnitude of the magnetic moment
measures the number of Schwinger bosons bound to the vortex and
determines the statistics of the vortex.

   The bond structure at the vortex core depends on where the
center of vortex is being located. Here we shall
consider only vortices which have maximum rotation symmetry allowed
by the underlying lattice, i.e. rotation by $n\pi/2$. In this case
there are only two possible type of vortices, with vortex center
located (i) at center of a plaquette and (ii) at a lattice point.
(see Fig.1) We shall first consider vortex (i).

   From continuum theory it is expected that the four bonds surrounding
the center of vortex (see Fig.1a) will be largely suppressed
because of the kinetic energy associated with the $V_{\mu}V^{\mu}$
term. Following similar analysis as in previous section we
find that the suppressed bonds introduce in the continuum theory
an effective electric quadrapole structure surrounding the
vortex center. Notice that as in the case of single broken bond the
structure is symmetric with respect to the $A$- and $B$- sublattices.
Thus we expect that localized spinon pairs may form around the
center of vortex but isolated magnetic moments cannot occur. Thus
this kind of vortex carries zero angular momentum and is a boson.

   The second kind of vortex is centered at a lattice point and
as a result, it is expected that the four bonds joining to the
vortex center (see Fig.1b) will be largely suppressed. This
situation is very similar to the case of single non-magnetic impurity
discussed in the last section. First of all, $2S$ spinons will be
found localized at the vortex center because of the suppressed
bonds. The spinons behaves as a 'staggered' gauge charge with
magnitide $2S$ and $2S$ other spinons with opposite
gauge charge will be nucleated from the vacuum to screen
the 'staggered' electric field generated by the spinons
localized at the vortex center. The spinons on opposite sublattices
interact antiferromagnetically through an effective Heisenberg
exchange of order $J'\sim{J}|\Delta_{c}|^{2}$, where $\Delta_{c}$ is
the bond amplitude of the four bonds joining to the vortex center.
As a result, a spin singlet will be formed. The vortex carries
zero angular momentum and is again a boson. Notice that this is
true even in the limit $J'\rightarrow0$ since integer number of bosons
$=4S$ are found binding to the vortex.

   The situation is however quite different if we consider a
finite density of the second kind of vortices. We shall show that
when density of vortices is large enough, it may become energetically
unfavourable to nucleate spinons from vacuum to screen the staggered
gauge field and as a result, only $2S$ spinons will be found
binding to vortex center resulting in fermionic vortices when $S$ is
half-integer (odd number of bound bosons). For simplicity we
shall first consider two vortices seperated by distance $l$, with
one vortex on each sublattice. First let us consider the energy when
spinons are nucleated from the vacuum to screen the staggered
gauge charge on each vortex. The total energy will be sum of
three terms: (i) the electrostatic energy which is of order
$2(2Se)^{2}ln(l_{o}/\xi)$, where $l_{o}\sim\xi$ is the 'size' of the
nucleated spinon wavefunction, (ii) The exchange energy between the
spinons localized as vortex center and nucleated spinons, which is
of order $-2S(S+1)J'e^{-l_{o}/\xi}\sim{-2S(S+1)J'}$,
and (iii) The energy needed to nucleate the
spinons from the vacuum, which is of order $4Sm$. The sum of
the three terms is of order $2S(2m-(S+1)J')$. The energy for the
second case when no spinons are nucleated from vacuum consists of
two terms: (i) the electrostatics energy which is of order
$(2Se)^{2}ln(l/\xi)$ and (ii)the exchange energy between spinons
located at the center of the two vortices which is or order
$-S(S+1)J'e^{-l/\xi}$.  In 2D, $e^{2}\sim{m}$ and the
electrostatic energy is of order $4S^{2}m.{ln}(l/\xi)$. For $l>>\xi$,
it is certainly energtically more favourable to nucleate bosons from
vacuum to screen the vortex gauge charge. However, when $l\leq\xi$,
the energy cost of the second case is of order $-S(S+1)J'$, and is
energetically more favourable if $S(S+1)J'<4Sm$. Notice that $J'
\sim{J}|\Delta_{c}|^{2}$ is expected to be very small because of
suppresion effect around vortex core. Thus the condition
can be satisfied easily even with a relatively small
$m$. For finite density of vortices $\delta$, $l\sim
\delta^{1 \over 2}$. Thus for low concentration of vortices, we
expect that spinons will be nucleated from vacuum to screen the
staggered gauge field and the second kind of vortices are bosons.
However, when density is high enough ($\delta\sim (1/\xi)^{2}$)
and $m$ large enough, the situation will change
and it may be energetically more favourable for
the vortices to stay as fermions (for half-integer $S$).
Notice that finite density analysis for the first kind of vortices
does not show similar qualitative difference between low and high
density. Because of the symmetry between $A$- and $B$- sublattice,
bosons are always added or removed from a single vortex in {\em
pairs} which cannot affect the statistics of a vortex.

   Before ending this section let's summarize our findings here.
Within a continuum description we have demonstrated in SBMFT
the existence of vortex solutions which are stable topological
excitations. We have found two different kind of vortices, according
to the different locations of vortex center. The first kind of
vortex has center located at center of a plaquette whereas the
second kind of vortex has center located at a lattice point.
The two kind of vortices have very different characters. The first
kind of vortex is always a boson, whereas the second kind of
vortex can be a fermion (for half-integer $S$) under conditions
which are rather easy to acheive. We cannot, however,
be absolutely sure whether 'fermionic' vortices exist
because of the natural limitation of a continuum theory which
gives only order of magnitude estimates. Notice also one important
distinction between the vortices we study in this paper and usual
vortices in superfluids. In the present case, vortices cannot
be distingished by sign of vorticity since
vortices carrying $1/2$ (or $\pi$) and $-1/2$ flux quanta
should be considered as identical. On a lattice where the
order parameter field $\Delta_{i,i+\mu}$ is a link-variable, $\pi$
and $-\pi$ vortices can be related by a pure gauge transformation.
The usual singularity encountered in the gauge transformation in
continous space does not arise here because of the discretized
lattice structure.

   It is important to clarify the different roles played by
the 'uniform' and 'staggered' gauge fields in
deciding the properties of vortices.
The gauge fields arise from fluctuation in phases of the order
parameter $\Delta_{i,i+\mu}$'s. The uniform gauge field couples to
bosons on the two sublattices with same gauge charge, and the
corresponding gauge symmetry is broken in SBMFT. Vortices
are stable because of this broken gauge symmetry, and corresponds to
solutions with "uniform" magnetic flux penetrating center of
vortices. The 'staggered' gauge field couples to bosons on opposite
sublattices with opposite gauge charge. It plays no role in
generating a stable vortex solution, but has strong effects in
determining the precise properties of the vortex core. Without the
'staggered' gauge field, spinons need not be nucleated from
vacuum to screen the 'staggered' charges generated at, for example,
the center of the second kind of vortex. In this case, for half-integer
spin systems, the second kind of vortices will always be
{\em fermions} independent of density, since only {\em odd} number
of bosons are found binding to center of each vortex (corresponding
to the spin $S$ at center of vortex). Similarly,
local magnetic moments will not be generated by non-magnetic
impurities if 'staggered' gauge field does not exist.

   It is also interesting to point out that similar vortex
excitations have been considered by Read and Chakraborty\cite{rc}
in a short-ranged RVB wavefunction for $S=1/2$ quantum
antiferromagnets. They considered also the two kinds of vortices
we discussed here. The statistics of the vortices were examined
by direct Berry phase computations where similar conclusion
that the first kind of vortex is a boson, whereas the
second kind of vortex is fermion independent of density is arrived
at their study. The effect of 'staggered' gauge field on vortices
is not included in their study. The vortices we discuss in
the present paper can be viewed as generalization of
their vortices in short-ranged RVB wavefunction to the more
complicated situation described by SBMFT.

\section{Quantum Liquid of Fermionic Vortices}
   In this section we shall examine the properties of a quantum liquid
of fermionic vortices based on general symmetry considerations.
We shall not ask the dynamical question of how the liquid of vortices
come to exist in the first place but shall ask the simpler question
if the liquid exists, what general properties should it has?
For simplicity we shall restrict ourselves to the case of
$S=1/2$ antiferromagnets which is the physical case of interests.
In this case, a unit fermionic vortex carries spin
$\pm{1/2}$. Notice that vorticity does not contribute another
independent quantum number as discussed in last section.

   First of all we shall consider the Fock space for vortices.
The 'vacuum' state for vortices is just a spin-disordered state in
SBMFT. A one vortex state with spin $\sigma$ at position $i$
is a vortex centered at site $i$ with spin pointing in direction
$\sigma$. Notice that two vortices with same $\sigma$ seperated
by distance $d$ are not orthogonal to each other but have finite
overlap $<\vec{x}_{1}|\vec{x}_{2}>\sim{e}^{-d/\xi}$. To construct
orthogonal set of vortex states, we consider Wannier vortex states
which can be constructed from vortex states using standard methods.
The Wannier states are orthonormal to each other and represent
localized vortices given that the overlap
$<\vec{x}_{1}|\vec{x}_{2}>$ falls off
exponentially as function of distance.\cite{rc}

   Next we shall introduce second quantization representation
for vortices. The creation operator $c^{+}_{i\sigma}$ is defined as
an operator which turns on an 'uniform' magnetic flux of $1/2$
flux quantum centered on site $i$, and creates a Schwinger boson
with spin $\sigma$ on site $i$. Similarly, a destruction operator
$c_{i\sigma}$ turns on an 'uniform' magnetic flux of $1/2$ flux
quantum in opposite direction, and removes a Schwinger boson with
spin $\sigma$ on site $i$. With the constraint that we cannot
put more than one Schwinger boson on any site, it can be checked
easily that these operators obey usual fermion commutation rules.
Notice that because of the constraint that there are always one
Schwinger boson per site, a {\em physical} vortex cannot be created
by $c^{+}_{i\sigma}$ operating on the 'vacuum' state, which changes
the number of Schwinger bosons on site $i$. An example of an
operator which creates a 'physical' vortex state from vacuum is
$c^{+}_{i\sigma}Z_{i\sigma}$, where $Z_{i\sigma}$ is a Schwinger
boson destruction operator with spin $\sigma$ on site $i$.
Notice also that the vortex occupation number $<n^{v}_{i\sigma}>
=<c^{+}_{i\sigma}c_{i\sigma}>$ is equal to the corresponding
Schwinger boson occupation number
$<n^{b}_{i\sigma}>=<\bar{Z}_{i\sigma}Z_{i\sigma}>$,
since the only difference between vortex and boson operators is
the magnetic flux which is being cancelled in the occupation
number operator.

   To proceed further we argue that the dynamics governing
the quantum liquid of vortices must possess a local $SU(2)$
symmetry, where the Lagrangian for vortices must be invariant
under the transformation

\begin{eqnarray}
c^{+}_{i\uparrow}\rightarrow\alpha_{i}c^{+}_{i\uparrow}+\beta_{i}
c_{i\downarrow}, \nonumber \\
c_{i\downarrow}\rightarrow{-\beta^{*}_{i}}c^{+}_{i\uparrow}+
\alpha^{*}_{i}c_{i\downarrow},
\label{su2}
\end{eqnarray}
with $|\alpha_{i}|^{2}+|\beta_{i}|^{2}=1$. Notice that the same $SU(2)$
symmetry was obtained in the Heisenberg model for $S=1/2$ spins when
the spins are expressed in terms of Fermion operators.\cite{aa,zr}

   The reason why we expect the local $SU(2)$ symmetry to exist can
be seen most easily by considering a physical vortex state with spin
$\sigma$ on site $i$ and ask the question: what is the
corresponding 'hole' or 'antiparticle' state for this vortex state?
To construct the 'hole' state we operate $c_{i\sigma}$ on the
'vacuum' state, which creates a magnetic flux of $1/2$ flux quantum
centered at site $i$, and remove a spin $\sigma$ Schwinger boson,
leaving site $i$ empty. However, this is not a physically allowable
state since the physically allowable state must has one boson per
site. To construct a physical 'hole' state we must put back a
Schwinger boson on site $i$. Now we may ask, which Schwinger boson
shall we put back? It cannot be a spin $\sigma$ Schwinger boson,
since putting back a spin $\sigma$ Schwinger boson just gives
the original 'particle' vortex state.  The only admissible choice
is thus to put on site $i$ a spin $-\sigma$ boson, which creates
a spin $-\sigma$ 'particle' vortex state. Thus we expect that
the 'hole' state for a spin $\sigma$ vortex is a spin $-\sigma$
vortex, which gives rise to the local $SU(2)$ symmetry stated
above. Another related argument is to observe that our underlying
system is a pure spin system. Thus the effective vortex
Hamiltonian must be equivalent to some effective pure {\em spin}
Hamiltonian. The spin operator when expressed in terms of fermions
possesses the $SU(2)$ symmetry stated above\cite{aa,zr}, so must
be a pure spin Hamiltonian.

    The $SU(2)$ local gauge invariance imposes a very severe
constraint on the plausible form of the effective theory for
the quantum liquid of vortices.  A mean-field theory of a quantum
liquid which is a spin singlet must describe a gas of fermions
with {\em particle-hole} symmetry, i.e., a half-filled
band with a particle-hole symmetric band-structure.
Fluctuations around the mean-field theory must be described by
a $SU(2)$ gauge theory. Physical excitations are represented by
particle-hole excitations (generation of vortex pairs).
These properties are all basic ingredients of the
slave-boson-mean-field theory at half-filling, when holes are
absent.\cite{aa,zr} Notice in particular that the mean-field
theory predicts that the Fermi surface for the vortices, if
exists, must be 'large', i.e., it obeys the Luttinger theorem
at half-filling, since $<n^{v}_{i\sigma}>=<n^{b}_{i\sigma}>
=1/2$ for spin-singlet wavefunctions.

   In the presence of holes the discussion is similar if one
assumes that holes form bound state with vortices. Notice that
in SBMFT where spins are bosons, holes are fermions and can be
represented by 'slave-fermion' operators $f^{+}_{i}$ and $f_{i}$.\cite{g5}
A hole binding to a spin $\sigma$ vortex is formed by removing
a spin $\sigma$ electron from the center of the spin $\sigma$
vortex. The resulting object is a {\em boson}, since it is
formed by binding a fermion (hole) to a magnetic flux of $1/2$
flux quantum. As in the case of vortices, we may introduce
boson creation and destruction operators $b^{+}_{i}$ and $b_{i}$
for the vortex-hole bound states. A boson creation operator
$b^{+}_{i}$ inserts a magnetic flux centered at site $i$, and
create a (fermionic) hole on site $i$. Notice that as in the case of
vortex operators, a {\em physical} hole cannot be created by
operating $b^{+}_{i}$ on the vacuum alone without first removing
the spin on site $i$. An example of an operator which creates a
physical vortex-hole bound state from vacuum is thus
$b^{+}_{i}Z_{i\sigma}$, where the spin on site $i$
is removed by the $Z$ operator. Notice also that the
vortex-hole bound state occupation number $<b^{+}_{i}b_{i}>$
is equal to the corresponding fermionic hole occupation number
$<f^{+}_{i}f_{i}>$, since the only difference between the
vortex-hole bound state operator and hole operator is the
appearance of the magnetic flux, which is again cancelled in
the occupation number operator.

   In the presence of holes the $SU(2)$ symmetry associated with
vortices will be broken. To see that we repeat our construction
of the 'hole' state for a spin $\sigma$ vortex. In presence of
holes the state generated by $c_{i\sigma}$ operating on the
vacuum is a physical state, since now holes are allowed in the
system. Thus the 'hole' state for a spin $\sigma$ vortex is
not necessarily a spin $-\sigma$ vortex, and the $SU(2)$ symmetry
is broken. The only symmetry which remains is the $U(1)$ symmetry
$c_{i\sigma}\rightarrow{c}_{i\sigma}e^{i\theta}, b^{+}_{i}
\rightarrow{b}^{+}_{i}e^{-i\theta}$, which follows from the
requirement that a physical electron destruction operator is
the product of the spin-annihilation operator and hole creation
operator and that all physical processes must be expressed in terms
of electron operators. Notice that because of the no double occupancy
constraint $<n^{v}_{i\uparrow}>+<n^{v}_{i\downarrow}>+
<b^{+}_{i}b_{i}>=1$, a mean field theory which treats the vortices
and vortex-hole bound states as independent gases of fermions and
bosons, respectively will produce a vortex fermi surface (if exists)
obeying the Luttinger theorem $<n^{v}_{i\uparrow}>=
<n^{v}_{i\downarrow}>=(1-\delta)/2$, where $\delta$ is the
concentration of holes in the system, as in the slave-boson mean
field theory for the $t-J$ model\cite{sb}.

\section{conclusion}
    In this paper we study the properties of vortices in disordered
state of SBMFT. Based on the continuum equations we show that
vortices are stable topological excitations in the system.
Furthermore, we show that because of the underlying lattice
structure, there exists two kind of vortices with vortex center
located at (i)center of a plaquette, and (ii)lattice point. The
two kind of vortices have different properties because of the
different symmetry with respect to the two sublattices. In
particular, for half-integer spin systems, we argue that the
second kind of vortices can be fermions, under conditions
which are not difficult to achieve. We next consider a quantum
liquid of fermionic vortices and show that because of general
symmetry requirement, the effective theory for the liquid of
vortices must has a general form very similar to
the slave-boson mean field theory in the undoped limit. In
the presence of holes similar conclusion is reached if one
assumes that holes form bound states with vortices. A very
interesting possibility brought out by our analysis is that
the Schwinger-boson-slave-fermion mean field theory which is
believed to describe the $t-J$ model qualitatively correctly
in the $\delta\rightarrow0$ limit and the slave-boson mean
field theory which is believed to describe the metallic state
more correctly can be understood within a single framework, if
one identify the fermionic spins in the slave-boson mean field
theory as vortices in Schwinger-boson treatment. In this paper
we have demonstrate the possibility of such an identification.
However, we have not been able to show within the framework
of Schwinger-boson-slave-fermion mean field theory that
fermionic vortices and vortex-hole bound states are
more important low energy excitations then bosonic spinwaves
and fermionic holes when density of holes is high enough.
The problem is related to the more fundamental question of
how the antiferromagnetic spin background get disordered when
holes are being introduced. We shall not try to address these
questions further in this paper, except to point out lastly that
based on the slave-fermion mean field theory for the $t-J$ model,
fermionic vortices and vortex-hole bound states indeed
become more stable in the presence of holes.

    In the presence of holes, the effective Hamiltonian for the
spins (in Schwinger-boson-slave-fermion treatment) has a form\cite{g5}

\begin{eqnarray}
H & = & -J\sum_{i,\eta}\left\{\Delta^{*}
\left[Z^{A}_{i\uparrow}Z^{B}_{i+\eta\downarrow}
-Z^{A}_{i\downarrow}Z^{B}_{i+\eta\uparrow}\right] + H.C.\right\}
\nonumber \\
  &   & +\lambda\sum_{i\alpha}\left[\sum_{\sigma}\bar{Z}^{\alpha}_{i\sigma}
Z^{\alpha}_{i\sigma}-(1-\delta)\right] \nonumber \\
  &   & +\sum_{i\sigma,\eta}\left\{(JQ^{*}+tF^{*})\bar{Z}^{A}_{i\sigma}
Z^{B}_{i+\eta\sigma}+ H.C. \right\}
\label{sf}
\end{eqnarray}
where $\eta=\pm\hat{x},\pm\hat{y}, \alpha=A,B,
 Q=\sum_{\sigma}<\bar{Z}^{A}_{i\sigma}
Z^{B}_{i+\eta\sigma}>$ and $F=<f^{+}_{i}f_{i+\eta}>$. The first term
is the original Heisenberg interaction and the $tF+JQ$ term is
an effective hopping term for spins generated by motion of holes.
The second term represents the constraint $\bar{Z}_{i}Z_{i}+
f^{+}_{i}f_{i}=1$ satisfied on average. Notice that the effective
Hamiltonian for Schwinger bosons still has a form very similar
to the BCS Hamiltonian for superconductors. Thus we expect
stable vortex solutions will exist. The most important
difference in the presence of holes is the existence of the
effective hopping term which breaks the {\em staggered gauge
symmetry}.\cite{ln} The 'staggered' gauge charge is no longer a
conserved quantity and as a result, the 'staggered' gauge field we
discuss in the previous sections will no longer be important\cite{ln}.
In particular, the second kind of vortex will be more likely to stay
as fermions since the 'staggered' electric field which is responsible
for binding another boson to the vortex is no longer important.
It is also easy to see that provided that fermionic vortices
exist, it is energetically favourable to bind holes to them. The
reason is simple: the antiferromagnetic correlation around the
vortex core is suppressed by the $V_{\mu}V^{\mu}$ term and
as a result, a hole can gain more kinetic energy by staying at
the center of a vortex.

   This work is supported by Hong Kong UGC Grant HKUST636/94P.

\appendix
\section{ }
   In this appendix we outline our derivation leading to the effective
Lagrangian Eq.\ (\ref{gl}). First we shall neglect
the staggered gauge field $A_{\mu}$ and shall set the terms $\phi(x)$,
$q_{\mu}(x)$ and $V_{\mu}(x)$ equal to constants, i.e. we neglect
gradient effects associated with these terms. In this case the
effective Lagrangian Eq.\  (\ref{lancon}) can be diagonalized
easily leading to Free energy (at zero temperature)

\begin{equation}
F = 2\sum_{\vec{k}}E(\vec{k}) + {1 \over 2}(2\phi^{2}+\vec{q}.\vec{q})
 -2(2S+1)(2\phi(1-{1 \over 8}\vec{V}.\vec{V})+m^{2})
\label{a1}
\end{equation}
where $E(\vec{k})=+(2\phi^{2}\vec{k}.\vec{k}-(\vec{q}.\vec{k})^{2}+
m^{2})^{1/2}$ is the energy for the (bosonic) spinons, $m$ is the spin
gap. The factor $2$ is coming from spin sum. In the limit
$m\rightarrow0$ and $\vec{q}\rightarrow0$,
$E(\vec{k})\rightarrow\surd{2}\phi|\vec{k}|$ and $F\rightarrow
(a-4(2S+1))\phi+\phi^{2}+(S+1/2)\phi(V_{\mu}V^{\mu})$, where $a$
is a cutoff dependent constant coming from $\sum{E}(\vec{k})$.
For $\vec{q}$ small, we can expand the $(\vec{q}.\vec{k})^{2}$
term in $E(\vec{k})$ to obtain the $q_{\mu}q^{\mu}$ term in Eq.\
(\ref{gl}). Notice that $\phi$ is always a postive real number since
the postive root is choosen in computing $E(\vec{k})$. Notice
also that the stability criteria $E(\vec{k})\geq0$ for all values of
$\vec{k}$ leads to the condition $\vec{q}.\vec{q}\leq{2}\phi^{2}$.
The terms involving the staggered gauge field $F_{\mu\nu}$ in
Eq.\ (\ref{gl}) can be obtained by
considering Gaussian fluctuations associated with the
$A_{\mu}$ field. Detailed calculation can be found in reference\cite
{rs} and we shall not repeat the derivation here.

\begin{figure}
\caption{two possible kind of vortices with vortex center located
 at (a) center of a plaquette, and (b)lattice point. In (a), the
 four bonds surrounding the vortex center and in (b), the four bonds
 joining to the vortex center, are suppressed.}
\label{Fig.1}
\end{figure}

\begin{references}
\bibitem{bd} The existence of bound state for arbitrary weak
 attractive potential can be seen in continuum theory by directly
 solving the (continuum) mean-field eigenvalue equation $(-\nabla^{2}
 +m^{2})\psi(\vec{x}) - V_{o}\delta(\vec{x})\psi(\vec{x}) = E^{2}
 \psi(\vec{x})$ for arbitrary small $V_{o}$.
\end{references}
\end{document}